\documentclass[
prb,
twocolumn,
superscriptaddress,
amsmath,amssymb,
aps,
]{revtex4-2}
\usepackage{appendix} 
\usepackage{graphicx}
\usepackage{dcolumn}
\usepackage{bm}
\usepackage[mathlines]{lineno}
\usepackage{braket}
\usepackage[T1]{fontenc}
\usepackage{comment}
\usepackage{xr-hyper}
\usepackage[ colorlinks, linkcolor=blue, anchorcolor=blue, citecolor=blue ]{hyperref}
\usepackage{soul,color,xcolor}
\makeatletter

\newcommand*{\addFileDependency}[1]{
\typeout{(#1)}
\@addtofilelist{#1}
\IfFileExists{#1}{}{\typeout{No file #1.}}
}\makeatother

\newcommand*{\myexternaldocument}[1]{%
\externaldocument[supp-]{#1}%
\addFileDependency{#1.tex}%
\addFileDependency{#1.aux}%
} 

\myexternaldocument{supporting}

\begin{document}
\title{Near room temperature magnetoelectric response and tunable magnetic anisotropy in the two-dimensional magnet 1T-$\rm CrTe_{2}$}
\author{Fengping Li}
\email{lifengping0109@gmail.com}
\affiliation{Department of Physics, University of Seoul, Seoul 02504, Korea}
\author{Bheema Lingam Chittari}
\affiliation{Department of Physical Sciences, Indian Institute of Science Education and Research Kolkata, Mohanpur 741246, West Bengal, India}
\author{Chao Lei}
\affiliation{Department of Physics, The University of Texas at Austin, Austin, Texas 78712, United States}
\author{Jeil Jung}
\email{jeiljung@uos.ac.kr}
\affiliation{Department of Physics, University of Seoul, Seoul 02504, Korea}

\date{\today}

\begin{abstract}
Magnets with controllable magnetization and high critical temperature are essential for practical spintronics devices, among which 
the two-dimensional 1T-$\rm CrTe_{2}$ stands out because of its high experimental critical temperature up to $\sim$300K down to the single layer limit. By using {\it ab initio} density functional theory, we investigate the magnetic properties of monolayer and bilayer 1T-$\rm CrTe_{2}$ and demonstrate that the magnetic properties, such as the magnetocrystalline anisotropy, critical Curie temperature and magnetizations, can be influenced by strain or electric fields.
\end{abstract}
\maketitle
\section{Introduction}\label{sec:level1}
The tunability of the magnetization in recently found two-dimensional (2D) magnets~\cite{2017Discovery,huang2017layer} by means of external electric and magnetic fields~\cite{faraday1846experimental,2011Electrical,2019Physical,2015Control,1998Magnetoelectronics,2000Electric,2010Multiferroic,chu2008electric} makes them interesting materials for potential spintronics applications~\cite{2004Spintronics,jungwirth2016antiferromagnetic}.
For instance, it has been shown that electrostatic doping can be used to adjust the magnetic characteristics of $\rm CrI_{3}$ layered antiferromagnetism ~\cite{2018Electrical,Jiang2018}, and that Dzyaloshinskii-Moriya interactions (DMI) enable the electric field to form topological magnon bands~\cite{2018Analysis}. 

Most proposed 2D magnets have demonstrated great tunability of their magnetic phases but their critical temperatures are well below room temperature, which restricts the practical applicability in devices. For instance, the critical temperatures of ferromagnetic (FM) monolayer and antiferromagnetic (AFM) bilayer $\rm CrI_{3}$ are all below 60~K \cite{huang2017layer}, and $\rm Cr_{2}Ge_{2}Te_{6}$ has a low Curie temperature of $\rm T_c$ of 66~K~\cite{gong2017discovery}. 
Higher $\rm T_c$ has been reported for the $\rm Fe_{3}GeTe_{2}$ ultrathin film Ising magnet, which is an itinerant ferromagnet with a Curie temperature of up to 130~K~\cite{wang2019current,fei2018two},
and several thin film dichalcogenides with FM or AFM 1-T polytypes ($\rm VSe_{2}$, $\rm VTe_{2}$, $\rm MnTe_{2}$, $\rm MnSe_{2}$, $\rm CrTe_{2}$) have been proposed as room temperature magnets~\cite{o2018room,bonilla2018strong,li2018synthesis,yu2019chemically,freitas2015ferromagnetism}.
Among these systems, ultrathin films of FM 1T-$\rm CrTe_{2}$ were reported to be layer antiferromagnets with in-plane magnetic anisotropy and have $\rm T_{c}$ close to room temperature, which have been verified experimentally for both bulk and thin films ~\cite{zhang2021room,meng2021anomalous,2021Ferromagnetism,purbawati2020plane,sun2020room,lasek2020molecular,2021Characterization,Gao2021}.
Recent theoretical works for 1T-$\rm CrTe_{2}$ have shown that the magnetic 
phase transition and charge density wave of the FM phase are strain tunable in monolayer 1T-$\rm CrTe_{2}$~\cite{lv2015strain,otero2020controlled,xian2022spin},
and theoretical predictions include anomalous Hall transport, magneto-optical effect and anomalous Nernst effect~\cite{yang2021tunable,li2021tunable}.

In this manuscript, we investigate the magnetism 
in monolayer and bilayer 1T-$\rm CrTe_{2}$ and explore the tunability of the magnetic properties subject to a perpendicular electric field and other system parameters like strains or Hubbard U repulsion. 
To this end we obtain by means of {\em ab initio} calculations the atomic orbital resolved electronic band structures and associated interatomic hopping terms, and assess the 
tunability of the magnetic anisotropy by means of strains, interlayer equilibrium distances and model choices for the onsite Hubbard U.
Total energies corresponding to different magnetic configurations allows to 
obtain the Hamiltonian parameters of the associated effective Heisenberg spin model, which is in turn used to study the critical temperatures for different magnetic anisotropies
and the magnetoelectric response of 
CrTe$_2$ bilayers in the presence of a perpendicular electric field. 

The manuscript is organized as follows. In Sec.~\ref{sec:methods} we introduce the theoretical methods employed in this paper. In Sec.~\ref{sec:electronic_magnetic} we present the electronic and magnetic properties of monolayer and bilayer 1T-$\rm CrTe_{2}$. In Sec.~\ref{sec:mae} we show the tunability of magnetocrystalline anisotropy energy due to strains. In Sec.~\ref{sec:magnetoelectric} we discuss the magnetoelectric response in bilayer 1T-$\rm CrTe_{2}$  under a perpendicular external electric field. In Sec.~\ref{sec:conclusion} we present the summary and conclusions. 

\section{Computational methods}\label{sec:methods}
The electronic structure, magnetic properties including the magnetic anisotropy, and exchange interactions are calculated by {\it ab initio} density-functional-theory (DFT)
as implemented in Quantum Espresso (QE)~\cite{2009Quantum}. 
For the exchange-correlation functional we used the Rappe Rabe Kaxiras Joannopoulos Ultrasoft (RRKJUS) Pseudopotentials for the Perdew-Burke-Ernzerhof (PBE) with the vdW-D2 correction to account for vdW interaction in 1T-$\rm CrTe_{2}$~\cite{perdew1996generalized,grimme2006semiempirical,barone2009role}. 
The cutoff energy for the plane-wave basis is set to 70~Ry and the kinetic energy cutoff for the charge density (potential) is set to 700~Ry, respectively. 
A vacuum region greater than 15~\AA  
~was used for monolayer and bilayer 1T-$\rm CrTe_{2}$ to avoid interaction between image layers along the $z$ direction. A 12 $\times$ 12 $\times$ 1 k-mesh was used to fully relax the triangular unit cell of monolayer and bilayer 1T-$\rm CrTe_{2}$. When calculating the exchange interactions, a rectangular supercell is constructed and a k-mesh of 9$\times$9$\times$1  and 6$\times$6$\times$1 is used for monolayer and bilayer 1T-$\rm CrTe_{2}$ respectively. The Monkhorst-Pack scheme is used for the k-mesh generation. 
The conjugate-gradient (CG) algorithm was adopted to relax the atomic positions until the maximum atomic force acting on the relaxed ions was better than $\rm 10^{-5}$~Ry/a.u., 
and an energy convergence criterion of $10^{-10}$~Ry was applied. 
The DFT+U approximation is considered for the correlation effects of Cr $3d$ electrons 
generally using U=2~eV, particularly for exploring the strain effects~\cite{anisimov1991band}
but we allow it to vary between 2 to 4~eV when exploring the electric field dependence of the bands. 
The structure and charge density distribution were visualized through the VESTA software~\cite{2011VESTA3}.

For the calculations of magnetic anisotropic energy, we 
carried out the static self-consistent calculations to obtain the wave function and charge density using the scalar-relativistic pseudopotential. 
The non-collinear calculations including the spin orbit coupling effect was performed using the previously obtained wavefunction and charge density by means of the fully-relativistic pseudopotential~\cite{chittari2020carrier,chittari2016electronic}. 
We use orbital projected bands representation and maximally localized Wannier functions to analyze the microscopic details of the magnetic states in monolayer and bilayer 1T-$\rm CrTe_{2}$ with a denser k-mesh of 12$\times$12$\times$1 
in the unit cell~\cite{Pizzi2020}. 
The in-plane biaxial strain was defined as $\varepsilon=\left(a-a_{0}\right) / a_{0} \times 100 \%$, where $a$ and $a_{0}$ are the strained and relaxed reference in-plane lattice constants.
The critical temperature and magnetoelectric response of 1T-$\rm CrTe_{2}$ layers 
were obtained numerically using the Metropolis Monte Carlo method~\cite{landau2021guide,newman2001monte}.

\section{DFT electronic and magnetic properties}
\label{sec:electronic_magnetic} 
Here we analyze the electronic and magnetic properties of single and bilayer 1T-$\rm CrTe_{2}$ by means of {\em ab initio} DFT calculations, where we assume that bilayer 1T-$\rm CrTe_{2}$ has the same AA stacking structure as the bulk crystal.
The bulk crystal structure of 1T-$\rm CrTe_{2}$ has a $P3m1$ group symmetry and 
an FM phase, and the measured lattice constants vary from 3.77 to 3.81\AA~ depending on experiments~\cite{freitas2015ferromagnetism,zhang2021room,meng2021anomalous}.
In the limit of single and bilayers our DFT calculations show that a monolayer 1T-$\rm CrTe_{2}$ has FM order while in the AA-stacked bilayer we have a layer AFM phase where the spins of each layer point in opposite directions~\cite{lv2015strain,li2021tunable,yang2021tunable} in keeping with earlier studies. 
In experiments, the bilayer 1T-$\mathrm{CrTe_2}$ can be exfoliated directly from the bulk crystal, which in turn has AA-stacking. Previous theoretical studies of bilayer systems have used lattice constants in the range of 3.7–3.9\AA~ to explore various magnetic orders and exchange interactions~\cite{yao2023control,wu2022plane,wang2020bethe,liu2022structural,li2021magnetic,katanin2025magnetic}.

In the following, we present a detailed analysis of intralayer FM in each CrTe$_2$ layer and interlayer AFM magnetization in a bilayer based on the tight-binding hopping terms between the $d$ orbitals of Cr atoms of the nonmagnetic phases.
In particular we analyze the $t_{2g}$-$t_{2g}$ and $t_{2g}$-$e_{g}$ hopping channels
based on the Kugel-Khomskii rule~\cite{kashin2020orbitally,mazurenko2006electronic,kugel1982jahn}: 
\begin{equation}
\label{khomskii}
J=-\frac{2(t^{t_{2g}-t_{2g}}_{1})^2}{U} +\frac{2(t^{t_{2g}-e_{g}}_{1})^2 J_{H}}{(U+\delta) (U+\delta-J_{H})}
\end{equation} 
to shed light in the magnetic exchange coupling processes in the system. 
At the same time, the effective exchange $J$ coupling parameters can be obtained from the total energy differences corresponding to different spin configurations.
For our calculations, we have used $U = 2$~eV for the Hubbard repulsion of $d$ electrons and $J_{H}\sim 0.4$~eV is the typical Hund's exchange constant for $3d$ Cr electrons~\cite{soriano2021environmental}. $\delta$ is the crystal field splitting energy between $t_{2g}$ and $e_{g}$ orbitals, that we can be estimated from the centroid of the corresponding projected density of states 
(see Fig. S1 which yields $\delta \approx $ 0.25 eV for monolayer and $\delta \approx $ 0.22 eV for bilayer\cite{susarla2021atomic}). Given its small value, we use $\delta = 0$ in our calculations.

The resulting exchange interaction values estimated from the NM phase's tight-binding hopping terms using Eq.~(\ref{khomskii}) are FM intralayer
$J^{\rm intra} = 0.38$~eV and AFM interlayer $J^{\rm inter} = -0.007$~eV 
respectively in monolayer and bilayer 1T-CrTe$_{2}$.
We will show that the signs agree qualitatively with the exchange coupling parameters obtained comparing DFT total energies of different magnetic states, but their magnitudes differ by almost an order of magnitude suggesting a high sensitivity of these parameters to calculation details. 
We further analyze the DFT magnetocrystalline anisotropy energies for single layer and AA-stacked bilayer and how they change depending on system parameters such as Hubbard U, in-plane strains and interlayer distances. 

\subsection{Intralayer FM and interlayer AFM in single and bilayer 1T-CrTe$_2$}
\begin{figure} 
\centering
\includegraphics[scale=0.66]{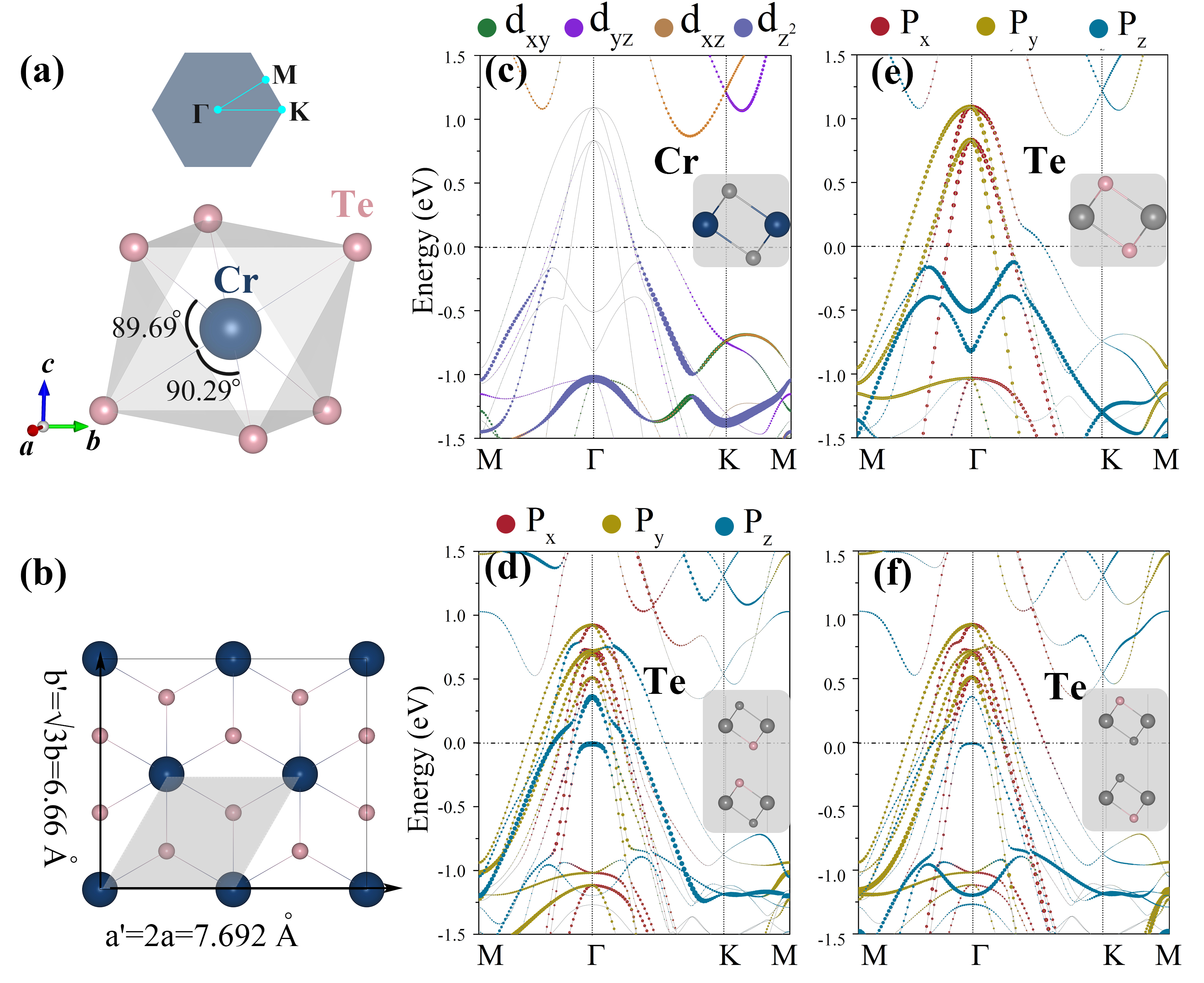}
\caption{\label{fig:structure} 
(a) Coordination structure and hexagonal Brillouin zone (BZ) of 1T-$\mathrm{CrTe}_{2}$. 
Blue and pink spheres represent Cr and Te atoms, respectively. 
(b) Schematic top view of the rectangular bilayer (BL) 1T-$\mathrm{CrTe}_{2}$ unit cell used for magnetic phases. 
The corresponding hexagonal unit cell is shown as a gray rhombus. 
(c) and (e) Orbital-projected band structures of monolayer 1T-$\mathrm{CrTe}_{2}$ for Cr $d$ orbitals and Te $p$ orbitals, respectively. 
Colored circles indicate the orbital contributions from Cr and Te atoms, where the color denotes orbital character and the circle radius is proportional to the orbital weight. 
(d) and (f) Orbital-projected band structures of bilayer 1T-$\mathrm{CrTe}_{2}$ for $p$ orbitals of interface Te atoms and outer Te atoms, respectively.
}
\end{figure}

Our DFT calculations predict that monolayer and bilayer 1T-$\rm CrTe_{2}$ have lowest energies for intralayer FM~\cite{lv2015strain,li2021tunable} and interlayer AFM~\cite{lv2015strain,li2021tunable,yang2021tunable} phases with optimized lattice constants of $a = 3.85\AA$ 
and $a = 3.86\AA$ respectively. 
The Fig.~\ref{fig:structure}(a) shows the relaxed crystal structure of monolayer 1T-$\rm CrTe_{2}$ 
giving rise to a triangular 2D layer of Cr atoms sandwiched between two triangular 2D layers of Te with C6 symmetry. 
The intralayer FM phase of monolayer 1T-$\rm CrTe_{2}$ whose Te-Cr-Te bond angle is close to $90^{\circ}$ is in keeping with the FM intralayer super-exchange coupling, analogously to CrI$_3$ monolayers.~\cite{sivadas2018stacking} 

\begin{table}[h!]
\caption{\label{tab:table3}
Total energies $\mathrm{E}_t$ (meV/atom) for different magnetic states of AA-stacked bilayer 1T-$\mathrm{CrTe}_{2}$. The energy of the lowest state is set to zero.}
\begin{ruledtabular}
\begin{tabular}{lcccc}
 & $\mathrm{BL}_{AFM}$ & $\mathrm{BL}_{FM}$ & $\mathrm{BL}_{Stripy}$ & $\mathrm{BL}_{Zigzag}$ \\
\hline
$\mathrm{E}_{t}$ & $0.0$ & $15.3$ & $7.5$ & $6.8$ \\
\end{tabular}
\end{ruledtabular}
\end{table}

We compared DFT total energies for different metastable spin configurations
to extract the effective exchange coupling parameters between the Cr atoms
using a rectangular supercell  containing 4~Cr atoms, as illustrated in Fig.~\ref{fig:structure}(b) to construct the FM, Stripy-AFM (Stripy), and Zigzag-AFM (Zigzag) spin patterns in a monolayer and in a bilayer sample. 
The total energies for these different spin configurations have been calculated using 
the same relaxed atomic structures of monolayer FM and intralayer AFM for bilayers. 
Examples of the magnetic configurations considered to obtain the exchange coupling parameters are shown in the supplementary {Fig.~S2}. We have verified that the ground states of monolayer prefer intralayer FM and bilayer 1T-$\rm CrTe_{2}$ prefer interlayer AFM for a range of Hubbard U values (U = 2/3/4 eV) based on the calculated total energies in different magnetic states, see Fig.~S3. 

In Table~\ref{tab:table3} we show the total energies corresponding to different magnetic phases in bilayer 1T-$\rm CrTe_{2}$ where an interlayer AFM coupling is energetically favored.
We defer to a later Sec.~\ref{sec:magnetoelectric} the discussion of the anisotropic Heisenberg model and associated magnetization based on these phenomenological exchange coupling parameters. 

In order to explore the microscopic mechanism of the intralayer FM and interlayer AFM phases, in the following we discuss the orbital content of the band structures and the tight-binding hopping parameters between the Cr atom orbitals in the NM phase. We find that the solutions have in-plane magnetic anisotropy and all calculations in our manuscript are presented for Hubbard U = 2~eV unless stated otherwise. 

We show in Fig.~\ref{fig:structure}(c)-(f) the spin polarized orbital projected electronic band structures of NM monolayer and bilayer 1T-$\rm CrTe_{2}$, and additional information is complemented in supplementary Fig.~S4 and Fig.~S5 for the spin-polarized band structures including the spin-orbit coupling terms, where we can see that the band structures around the Fermi level are mainly contributed by the $d$ orbitals of Cr atoms and the $p$ orbitals of Te atoms. The $d$-shell of Cr atoms split into two sets of inequivalent $t_{2g}$($d_{xz}$, $d_{yz}$,$d_{xy}$) and $e_{g}$ ($d_{z^{2}}$,$d_{x^{2}-y^{2}}$) orbitals due to the octahedral crystal fields. According to the DFT results, the Cr atoms have a magnetic moment of $3 \mu_B$ and thus have three unpaired electrons in the $t_{2g}$ orbitals. 
As a rule of thumb we posit that the hopping term between equal spin $t_{2g}$ electrons is suppressed due to the conservation of spin angular momentum and Pauli exclusion principle
while they are allowed when the spins point in opposite senses.
For equal spins only the hopping terms between $t_{2g}$ to $e_g$ orbitals are allowed.

Therefore, the degree of $t_{2g}$-$t_{2g}$ ($t_{2g}$-$e_{g}$) hybridization should favor the AFM (FM) exchange interaction between the Cr atoms. We estimated the hopping parameters for $t_{2g}$-$t_{2g}$, $t_{2g}$-$e_{g}$ orbitals between the adjacent intralayer and interlayer Cr atoms by considering the virtual hopping processes through the Te atoms.

\begin{figure}
\includegraphics[scale=0.7]{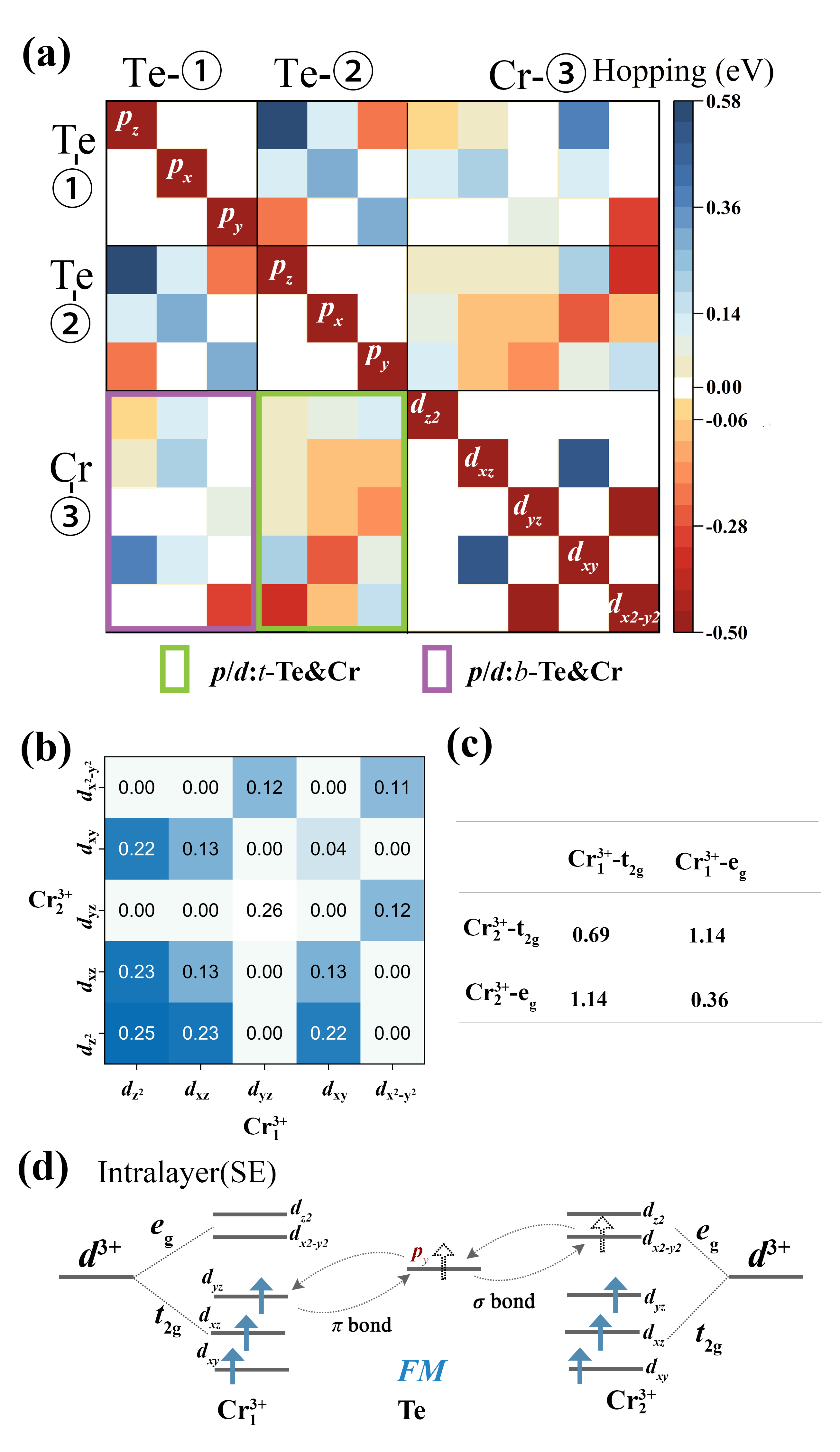}
\caption{\label{fig:orbitals} 
(a) Orbital-resolved hopping parameters for $p$ and $d$ orbitals in monolayer 1T-$\mathrm{CrTe}_{2}$. 
The green and purple rectangles represent $p$- and $d$-orbital hopping interactions between Te and Cr atoms, denoted as $t$-Te–Cr and $b$-Te–Cr, respectively. 
To clearly visualize the relative magnitudes of the hopping interactions, parameters close to zero are shown in white. 
(b) Hopping terms of $d$ orbitals between adjacent Cr atoms ($\mathrm{Cr}_{1}^{3+}$ and $\mathrm{Cr}_{2}^{3+}$). 
(c) Effective $d$-orbital hopping terms between adjacent $\mathrm{Cr}_{1}^{3+}$ and $\mathrm{Cr}_{2}^{3+}$ atoms, classified into $t_{2g}$ and $e_{g}$ symmetries. 
(d) Schematic illustration of the evolution of electronic states along two Cr–Te–Cr superexchange (SE) pathways, showing the ferromagnetic (FM) alignment of intralayer $\mathrm{Cr}^{3+}$ cations.
}
\end{figure}

\begin{figure*}
\centering
\includegraphics[scale=0.6]{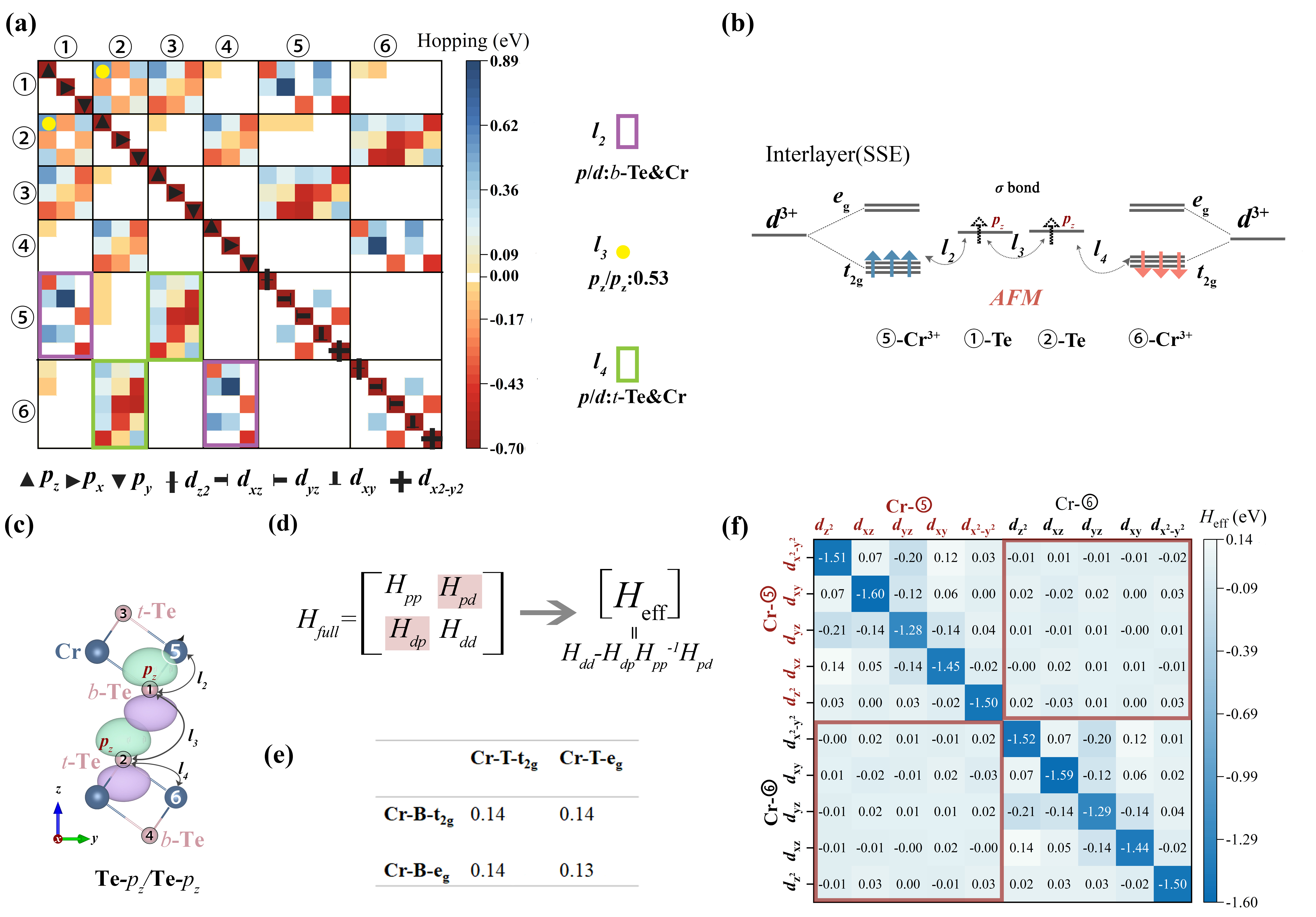}
\caption{\label{fig:hopping}
(a) Orbital-resolved hopping parameters for $p$ and $d$ orbitals in bilayer 1T-$\mathrm{CrTe}_{2}$. 
The atomic indices are indicated in Figs.~\ref{fig:hopping}(c). 
To clearly visualize the magnitudes of the hopping interactions, parameters close to zero are shown in white. 
The green and purple rectangles represent $p$- and $d$-orbital hopping interactions between top (t) and bottom (b) Te atoms and Cr atoms, denoted as $t$-Te and $b$-Te, respectively. 
(b) Schematic illustration of the evolution of electronic states along the Cr–Te–Te–Cr super-superexchange (SSE) pathway, showing the antiferromagnetic (AFM) alignment of interlayer $\mathrm{Cr}^{3+}$ cations. 
(c) Maximally localized Wannier functions (MLWFs) of interface Te $p_{z}$ orbitals in bilayer 1T-$\mathrm{CrTe}_{2}$. 
$l_{2}$, $l_{3}$, and $l_{4}$ denote the exchange interaction paths between atoms, corresponding to Fig.~\ref{fig:hopping}(b). 
(d) Transformation of $2\times 2$ hopping-term blocks into an effective hopping matrix using degenerate-state perturbation theory. 
(e) Effective hopping terms between top-layer and bottom-layer Cr atoms, classified into $t_{2g}$ and $e_{g}$ symmetries. 
Here, T and B denote the top and bottom layers, respectively. 
(f) Explicit effective hopping terms between top-layer and bottom-layer Cr atoms.}
\end{figure*}
For FM monolayer, we show the bands resulting from the $pd\pi$ electronic hybridization between $p_{z}$ of Te atom and $t_{2g}$ ($d_{yz}$ \& $d_{xz}$) orbitals of Cr atoms in Fig.~\ref{fig:structure}(c) and (e)~\cite{kim2019giant,khomskii2014transition}.  
The simplest picture for intralayer FM in 1T-CrTe$_2$ is the super-exchange between the magnetic Cr $3d$ and ligand Te $p$ orbitals in line with the Goodenough-Kanamori-Anderson (GKA) rules.
The magnitude of the hopping parameters between three $p$ orbitals of Te atom and five $d$ orbitals of Cr atom are illustrated in Fig.~\ref{fig:orbitals}(a) of monolayer 1T-$\rm CrTe_{2}$, and are highlighted with green and purple boxes while the diagonal elements correspond to on-site energies. The DFT hopping parameters between the five $d$ orbitals at the Cr sites are listed in Fig.~\ref{fig:orbitals}(b), and the effective hopping parameters between $t_{2g}$/$e_{g}$ orbitals are shown in Fig.~\ref{fig:orbitals}(c) where the $t_{2g}$-$e_g$ effective hopping term is about $\sim$456~meV larger than $t_{2g}$-$t_{2g}$ hopping term, justifying the favoring of intralayer FM in a monolayer~\cite{xiao2020modulating} based on the superexchange interaction illustrated in Fig.~\ref{fig:orbitals}(d). 

\begin{figure*}
\centering
\includegraphics[scale=0.27]{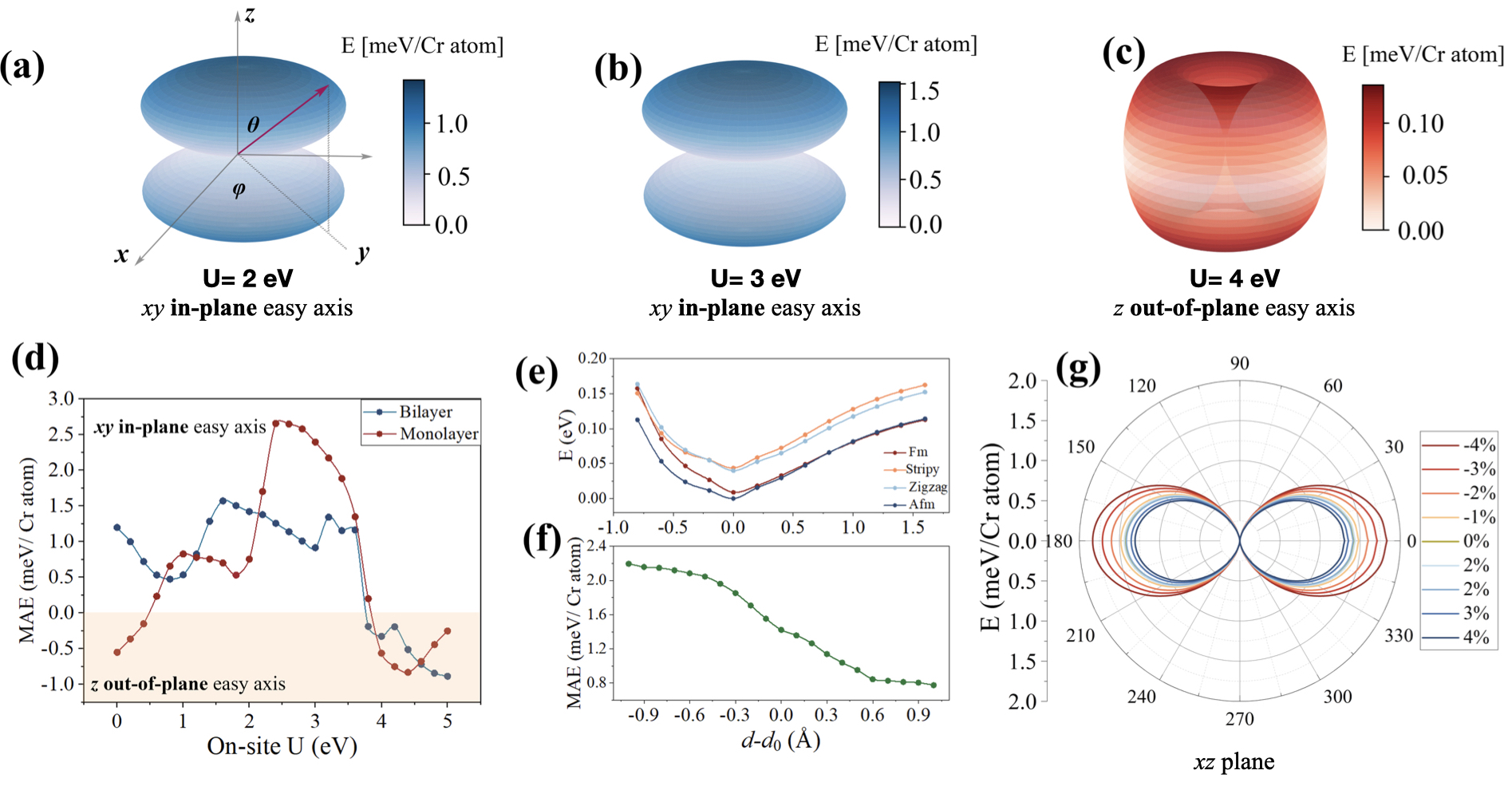}
\caption{\label{fig:magneticAnistropyEnergy} 
Magnetic anisotropy energy (MAE) in bilayer (BL) 1T-$\mathrm{CrTe}_{2}$ under different electronic correlations ($U$), interlayer distances and in-plane strains.
(a–c) Spherical plots of the ground-state energies for $U = 2$, $3$, and $4$ eV, respectively. 
Darker colors and larger radial distances correspond to higher MAE values. 
The spin orientation $\mathbf{\hat{s}}(\theta,\varphi)$ is indicated by a red arrow, where $\theta$ and $\varphi$ denote the azimuthal and polar angles, respectively. 
(d) MAE along the $xz$ plane as a function of on-site electron–electron interaction $U$ (0–5 eV). 
The red dotted line corresponds to the monolayer, and the blue dotted line corresponds to the bilayer 1T-$\mathrm{CrTe}_{2}$. 
(e) Ground-state energies of four different magnetic states as a function of interlayer Cr–Cr distance, with the lowest energy set as the reference. 
(f) MAE along the $xz$ plane versus interlayer Cr–Cr distance in bilayer 1T-$\mathrm{CrTe}_{2}$. 
(g) MAE as a function of in-plane strain ($-4\%$ to $+4\%$) in bilayer 1T-$\mathrm{CrTe}_{2}$ where spins are rotated in the $xz$ plane.
}
\end{figure*}

For bilayer 1T-$\rm CrTe_{2}$, the DFT calculations predict FM intralayer magnetism and interlayer AFM that originate from the super-super-exchange interaction via Cr-Te-Te-Cr atoms~\cite{lv2015strain,Gao2021}. 
Beyond the energetic analysis of the magnetic phase, our work provides a microscopic theory justification based on calculated orbital interactions, a perspective that has not previously been addressed.   
An analysis of the orbital projected bands in Fig.~\ref{fig:structure}(d) and \ref{fig:structure}(f) indicate that occupied electrons near the Fermi level that influence the magnetization locate preferentially at the $p_z$ orbitals of the inner interface Te atoms than the outer interface. Morover, the charge difference density in Fig.~S6 shows a large charge localization at the interface area, which also indicates stronger interface coupling interaction.

 Here we explore quantitatively the interlayer magnetic exchange interaction based on the super-super-exchange (SSE) interactions in bilayers as illustrated in Fig.~\ref{fig:hopping}(b). 
 The projected hopping parameters between $p$ and $d$ orbitals for six atoms in bilayer 1T-$\rm CrTe_{2}$ are shown in Fig.~\ref{fig:hopping}(a), together with the associated atom indices in Fig.~\ref{fig:hopping}(c). Similar to the monolayer case, the hopping parameters highlighted with purple and green rectangles in Fig.~\ref{fig:hopping}(a) indicate that the intralayer exchange interactions are ferromagnetic. The interlayer magnetic exchange interactions should stem mainly from SSE by hopping processes through Cr-Te-Te-Cr interlayer ligands. In order to estimate the SSE between the top and bottom $d$ orbitals of Cr atoms, we employ the second order perturbation theory to calculate the effective hopping term between $d$ orbitals of Cr atoms as illustrated in Fig.~\ref{fig:hopping}(d)~\cite{sakurai1995modern}.

The effective $t_{2g}$–$e_g$ hopping parameters shown in Fig.~\ref{fig:hopping}(e) are comparable to the $t_{2g}$–$t_{2g}$ values in Fig.~\ref{fig:hopping}(f), with the latter being $\sim$6~meV larger, thereby indicating a slight preference for the layer AFM state according to Eq.~(\ref{khomskii}). 
This trend is consistent with our DFT calculations. More broadly, the analysis of effective orbital interactions provides a complementary framework for predicting the magnetic phase of 1T-$\rm CrTe_2$, offering an alternative to conventional total-energy comparisons that can be used in other correlated magnetic materials.

\subsection{Tunable magnetocrystalline anisotropy  energy}\label{sec:mae}

The MAE stem from spin-orbit interactions and facilitates the formation of long-range FM order 
in the two-dimensional limit by aligning the magnetic moments along an easy direction.
Previous studies reveal that the magnetization in 1T-$\rm CrTe_2$ is highly tunable, depending on structural and external factors. In the monolayer, the easy axis is typically out-of-plane but can be switched in-plane by doping, strain, or intercalation~\cite{liu2022structural,yao2023control,wu2022plane}. Bilayer and multilayer systems generally favor in-plane alignment, sometimes forming noncollinear or zigzag textures influenced by stacking and substrate effects~\cite{li2021magnetic,purbawati2020plane,kashin2020orbitally}. A 120° in-plane antiferromagnetic order has also been theoretically predicted for the monolayer~\cite{katanin2025magnetic}, underscoring the strong tunability of magnetic anisotropy in 1T-$\rm CrTe_2$.
Here, we explore the sensitivity of the MAE to Hubbard U strengths and strains. 
The effects of Hubbard U are investigated in monolayer and bilayer 1T-$\rm CrTe_{2}$ with varied values (U=2/3/4 eV). 
Both the MAE and magnetic moments depend on the electronic correlations, the latter increasing in magnitude from $2.7 \mu_B$ to $3.7 \mu_B$ when U is altered from 0 to 5 eV, see Fig.~S7.

Firstly, we assess the impact of electronic correlations U on the MAE by obtaining the ground state energies versus the direction of magnetic moments in a bilayer 1T-CrTe$_2$ as illustrated in Fig.~\ref{fig:magneticAnistropyEnergy}(a)-(c)
and for a monolayer in Fig.~S8(a)-S8(c) where we use U = 2,3 and 4~eV. For both systems we see that the in-plane $xy$ easy axis points towards the out of plane $z$-axis for sufficiently large U value around U= 3.8~eV. 
Similar findings were also reported in the Co thin film study ~\cite{1998Perpendicular}.
The fact that various substrate screening alters the on-site electronic corrections and hopping interaction of $d$ orbitals on Cr atoms may provide an explanation for the discrepancy of MAE results reported in previous experiments~\cite{wang2024strain}.
Monolayer 1T-CrTe$_2$ grown on bilayer graphene exhibits near room-temperature ferromagnetism with an in-plane easy axis, showing a magnetic anisotropy energy (MAE) of $\approx 1.4$~meV per unit cell induced by $\sim 1.6\%$ epitaxial strain and electron doping~\cite{wang2024strain}. In bilayer CrTe$_2$, strain and intercalation tune the MAE from $-0.6$ to $+2$~meV per unit cell, switching the easy axis between in-plane and out-of-plane via modulation of Te~$5p$ orbitals~\cite{li2021magnetic}. These findings underscore the decisive influence of strain, carrier doping, and interlayer effects on magnetic anisotropy in 1T-CrTe$_2$.

The ground state energies of monolayer and bilayer 1T-$\rm CrTe_{2}$ with in-plane magnetic moment directions are also provided in Fig.~S9(a)-(c) and  Fig.~S9 (d)-(f), respectively. The in-plane magnetic anisotropy energies are almost three orders of magnitude smaller than the out-of-plane ones.
There are 6 in-plane easy directions according to the magnetic energy calculations, which is compatible with the $C_{3v}$ symmetry of 1T-$\rm CrTe_{2}$. In addition, the dependency of MAE on Hubbard U values shown in Fig.~\ref{fig:magneticAnistropyEnergy}(d) indicates that the easy direction of magnetic moments switches from out-of-plane to in-plane at the value of U= 3.8 eV.  

As previously mentioned the out-of-plane strain is an effective tool to regulate the magnetic states in bilayer van der Waals materials. To explore the strain effect, we adjust the interlayer distance of bilayer 1T-$\rm CrTe_{2}$ and contrast the ground state energies for various magnetic states~\cite{2021Magnetoelectric}, as shown in Fig. \ref{fig:magneticAnistropyEnergy}(e).  
The magnetic state is layer AFM when interlayer distance is at equilibrium, and becomes layer FM when the interlayer distance increases. 
 The interlayer magnetic coupling in bilayer 1T-$\rm CrTe_{2}$ is highly sensitive to the interlayer distance, following a Bethe–Slater-like behavior~\cite{li2021magnetic,wang2020bethe}. Increasing the Te–Te separation weakens Pauli repulsion and enhances kinetic exchange, driving an antiferromagnetic-to-ferromagnetic (AFM–FM) transition mediated by Te $p$-orbital superexchange.

As we discuss below, strain can be used to adjust the MAE. The MAE decreases when we increase the interlayer distance when the spin orientation variable points in the out-of-plane direction as seen in Fig.~\ref{fig:magneticAnistropyEnergy}(f). 
When the Cr–Cr interlayer distance increases by approximately 1~\AA, the MAE converges to about 0.75~meV per atom, indicating an effective decoupling of the bilayer into two independent monolayer 1T-$\rm CrTe_{2}$ sheets.
Inspired by experimental efforts utilizing various substrates, we further examine the evolution of the MAE under biaxial in-plane strain. The variations in ground-state energies and lattice parameters as a function of strain  ($\pm 4\%$) are presented in Fig.~\ref{fig:magneticAnistropyEnergy}(f) and further complemented in Figs.~S10(a) and S10(b).
Correspondingly, the strain-dependent magnetic configurations, shown in Figs. S10(c) and S10(d), reveal that the ferromagnetic (FM) and antiferromagnetic (AFM) states constitute the energetically most favorable configurations for the monolayer and bilayer systems, respectively.
MAE increases when applying biaxial strain both in monolayer see Fig.~\ref{fig:magneticAnistropyEnergy}(g) and bilayer 1T-$\rm CrTe_{2}$, see Fig.S11(a), and in-plane MAE shows $\phi = 30^\circ$ and $\phi = 60^\circ$ easy axis as shown in Fig.~S11 (b) and S11(c).

Overall we have shown the dependence of MAE on the strength of onsite U Coulomb correlations and its tunability through both out-of-plane and in-plane strains.

\begin{figure}
\centering
\includegraphics[scale=0.15]{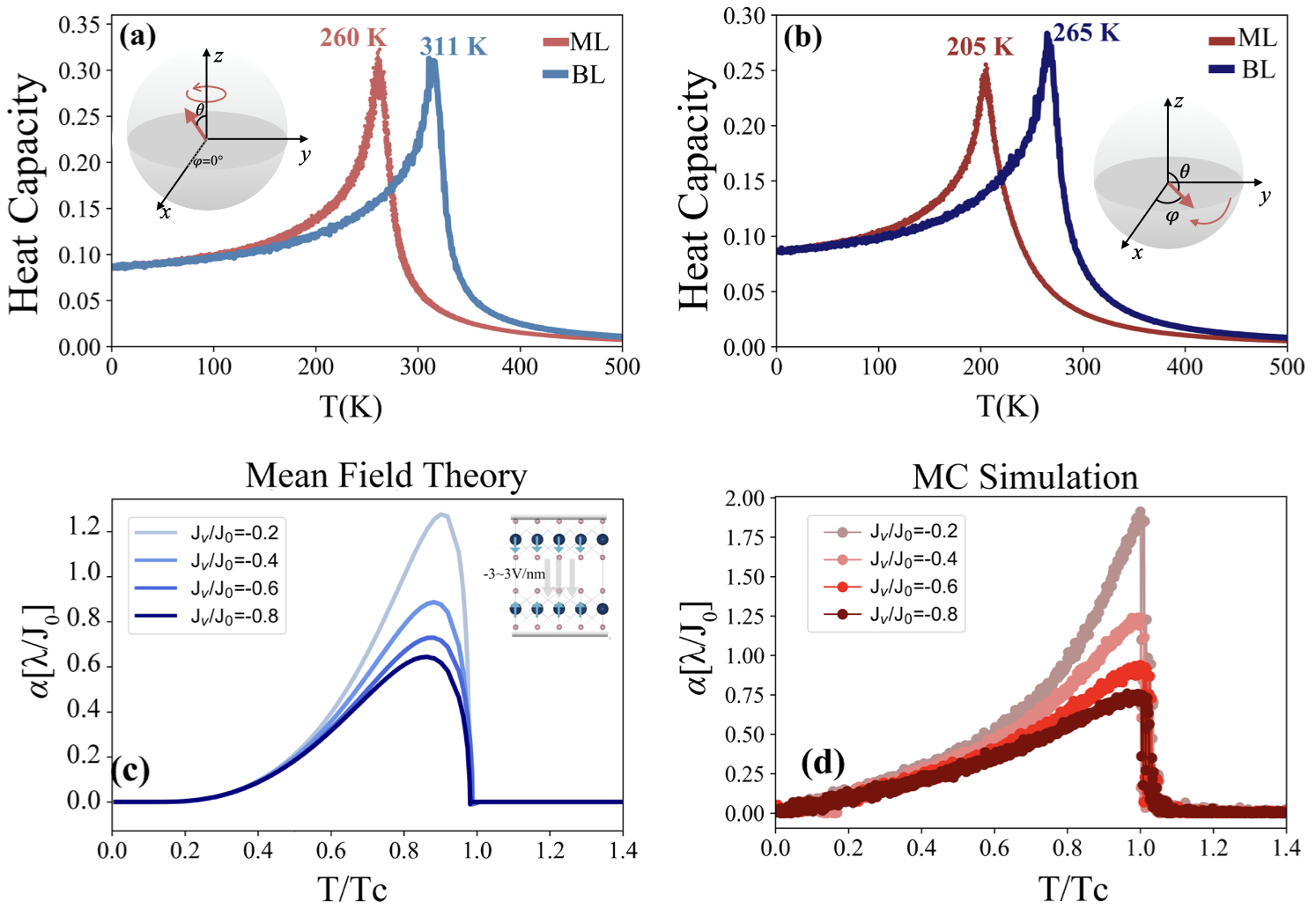}
\caption{\label{fig:temperature-magntoelectroni}
(a) Temperature dependence of the heat capacity in monolayer (ML) and bilayer (BL) 1T-$\mathrm{CrTe_{2}}$ considering only out-of-plane anisotropy (polar angle $\varphi = 90^{\circ}$, azimuthal angle $\theta$ variable). The definition follows Eq.~(\ref{maeo}), where $E_{xy}$ and $E_{z}$ denote the in-plane and out-of-plane anisotropy energies, respectively. (b) Temperature dependence of the heat capacity in ML and BL 1T-$\mathrm{CrTe_{2}}$ considering only in-plane anisotropy (azimuthal angle $\theta = 90^{\circ}$, polar angle $\varphi$ variable). The definition follows Eq.~(\ref{maei}), where $E_{x}$ and $E_{y}$ correspond to the anisotropy energies along the in-plane $x$ and $y$ directions, respectively. (c) Magnetoelectric response of bilayer 1T-$\mathrm{CrTe_{2}}$ obtained from mean-field theory and (d) numerical Monte Carlo simulations. (Inset of (c)) Schematic side view of a polarized bilayer 1T-$\mathrm{CrTe_{2}}$ in the AFM phase, illustrating the magnetoelectric response induced by an applied vertical electric field ($-3$ to $+3$ V/nm).}
\end{figure}

\section{Magnetoelectric response}
\label{sec:magnetoelectric}
In this section, we discuss the magnetoelectric effect of bilayer 1T-$\rm CrTe_{2}$ when we apply a perpendicular electric field, where we expect a behavior similar to the prediction for a CrI$_3$ bilayer~\cite{2021Magnetoelectric}. 
A strong layer magnetoelectric response appears near the critical temperature in these van der Waals bilayers thanks to a relatively weaker interlayer AF exchange coupling versus in-plane FM coupling.
As mentioned above and shown in Fig.~S2, the magnetic states are unaffected by electric fields as large as 3~eV/nm. However, the electric field does change the intralayer magnetic coupling strengths and therefore the temperature dependent response. 
We have followed a similar analysis to Ref.~\cite{2021Magnetoelectric} to find the critical temperature ($\rm T_{c}$) in  FM monolayer (Curie temperature) and AFM(N$\rm \acute{e}$el temperature) bilayer 1T-$\rm CrTe_{2}$ by means of Metropolis Monte-Carlo (MC) simulation, based on classical Heisenberg Hamiltonian triangle Cr-ion spin network:

\begin{equation}
H=-\frac{1}{2}\left(\sum_{<i j>} J_{1} \vec{S}_{i} \cdot \vec{S}_{j}+\sum_{\ll i j\gg} J_{2} \vec{S}_{i} \cdot \vec{S}_{j}\right)-J_{v} \sum_{<i j>} \vec{S}_{i} \cdot \vec{S}_{j},
\end{equation}
where $J_{1}$ and $J_{2}$ are the first nearest and second nearest neighbor intralayer exchange coupling, respectively. $J_{v}$ denotes the interlayer exchange coupling in bilayer 1T-$\rm CrTe_{2}$. $\vec{S}_{i}$ is the spin magnetic moment from the Cr atom whose position is labeled by the symbol $i$. Our $ab$ $initio$ calculations indicate that each Cr atom has the magnetic moment of around 3$\mu_{B}$, thus we have $S=M / g \mu_{B} = 3/2$.
In order to obtain  $J_{1}$, $J_{2}$ and $J_{v}$ in monolayer and bilayer 1T-$\rm CrTe_{2}$, the energies of various magnetic configurations in a rectangular supercell are calculated and mapped with the Heisenberg model, with the magnetic configurations shown in Fig. S1. The calculation details of the exchange coupling parameters are illustrated in the Supporting Information. 

\begin{table}[h!]
\caption{\label{tab:table2}
Intralayer and interlayer exchange coupling parameters for monolayer and bilayer 1T‑$\mathrm{CrTe_{2}}$. $\mathrm{E}_{O}$ and $\mathrm{E}_{I}$ (meV) are respectively the spin reorientation magnetic anisotropy energies from out‑of‑plane to in‑plane, and from in‑plane $x$‑ to $y$‑axis. $\gamma$ and $\Delta$ represent spin anisotropy parameters along the out‑of‑plane and in‑plane directions, and $\mathrm{J}_{v}$ (meV) is the interlayer exchange coupling obtained from total energy differences.}
\begin{ruledtabular}
\begin{tabular}{cccccccc}
& $\mathrm{J}_{1}$ & $\mathrm{~J}_{2}$ & $\mathrm{E}_O$ & $\mathrm{E}_{I}$ & $\boldsymbol{\gamma}$ & $\Delta$ & $\mathrm{~J}_{v}$\\
\hline $\mathrm{M L}$ & $7.45637$ & $3.43808$ & $4.251$ & $0.077$ & $0.6088$ & $0.498$ \\
$\mathrm{B L}$ & $8.21757$ & $2.62172$ & $4.473$ & $0.0199$ & $0.6055$ & $0.5099$ & $-6.72$
\end{tabular}
\end{ruledtabular}
\end{table}

In order to estimate the magnetoelectric response, we get the electric field dependence of the intralayer exchange coupling parameters, as a function of various Hubbard U corrections, as shown in Fig.~S12. Although the Hubbard U corrections enhance the absolute value of the exchange interactions, they have no impact on the electric field dependence. 
Similar to previous discussions, U = 2 eV is used for estimating the $T_{c}$ in monolayer and bilayer 1T-$\rm CrTe_{2}$ as shown in Fig.~S13. MAE needs also to be taken into account in the Heisenberg model in order to estimate the critical temperature $T_c$.  The following definitions apply to both out-of-plane and in-plane MAEs:
\begin{equation}
\text {MAE}_{O} = E_{O}=\mid E_{xy} - E_{z}\mid,
\label{maeo}
\end{equation}
where we chose $E_{xy} = E(\theta = \pi/2, \varphi =0^{\circ})$ in between the $x$ and $y$ axes, and the in-plane total energy difference
\begin{equation}
\mathrm{MAE}_{I} = E_{I} = \left|  E_{y}  - E_{x} \right|.
\label{maei}
\end{equation}
The total energy including the spin anisotropy $\boldsymbol{\gamma}$ from out-of-plane to in-plane x-axis is thus:
\begin{equation}
\begin{aligned}
E_{x y}=&-(1-\gamma)\left[\left(J_{1 x} S_{i}^{x}+J_{2 x} S_{i}^{x}\right) \sum_{j} s_{j}^{x}\right.\\
&\left.+\left(J_{1 y} S_{i}^{y}+J_{2 y} S_{i}^{y}\right) \sum_{j} s_{j}^{y}\right] \\
E_{z}=-& \gamma\left[\left(J_{1 z} S_{i}^{z}+J_{2 z} S_{i}^{z}\right) \sum_{j} s_{j}^{z}\right].
\end{aligned}
\end{equation}
Table \ref{tab:table2} provides a summary of the intralayer exchange coupling parameter, MAE and the resultant spin anisotropy. Based on these variables, we can determine the relationship between heat capacity and temperature in monolayer and bilayer 1T-$\rm CrTe_{2}$ as shown in Fig. \ref{fig:temperature-magntoelectroni}(a), indicates that $T_{c}$ are 260 K and 311 K, respectively.

When considering the in-plane MAE in monolayer and bilayer 1T-$\rm CrTe_{2}$, the total energy within $xy$ plane including spin anisotropy parameter can be expressed as follows: 
\begin{equation}
\begin{array}{l}
E_{x}=-\Delta \left[\left(J_{1 x} S_{i}^{x}+J_{2 x} S_{i}^{x}\right) \sum_{j} s_{j}^{x}\right] \\
E_{y}=-(1-\Delta)\left[\left(J_{1 y} S_{i}^{y}+J_{2 y} S_{i}^{y}\right) \sum_{j} s_{j}^{y}\right] \\
E_{z}=0
\end{array}
\end{equation}
and the calculated heat capacity as a function of temperature is shown in
Fig.~\ref{fig:temperature-magntoelectroni}(b).
The resulting $T_{c}$ is closer to experiments with values of 205~K for monolayer and 265~K for bilayer 1T-$\rm CrTe_{2}$~\cite{xian2022spin}. 
We estimated the magnetoelectric coefficient in bilayer 1T-$\rm CrTe_{2}$ under a perpendicular electric field through Monte-Carlo simulation. The exchange coupling parameters of the top and bottom layers in the presence of an electric field were obtained by DFT calculations.  Specifically, the $J_{1}$ and $J_{2}$ parameters in Eq.~(\ref{Ham_bilayer}) for top and bottom layers are estimated using DFT and shown in Fig.~S11(c) and S11(d).
The Hamiltonian of a bilayer van der Waals magnet is written as:
\begin{equation}\label{Ham_bilayer}
\begin{aligned}
H=&-J_{t 1} \sum_{<i j>} \vec{S}_{t i} \cdot \vec{S}_{t j}-J_{t 2} \sum_{\ll i j\gg} \vec{S}_{t i} \cdot \vec{S}_{t j} \\
&-J_{b 1} \sum_{<i j>} \vec{S}_{b i} \cdot \vec{S}_{b j}-J_{b 2} \sum_{\ll i j\gg} \vec{S}_{b i} \cdot \vec{S}_{b j} \\
&-J_{v} \sum_{<i j>} \vec{S}_{t i} \cdot \vec{S}_{b j}
\end{aligned}
\end{equation}
 In Monte Carlo simulations, the spin correlation function of  $\alpha=\left\langle S_{z} 
 \partial H / \partial E\right\rangle / (N k_{B} T)$
 is used to estimate the magnetoelectric coefficient, where $H$, $S_{z}$ and $N$ is the classical spin-Hamiltonian, total magnetization and the number of lattice sites per layer, respectively~\cite{2010Temperature}. 
The relevant interlayer and intralayer exchange coupling parameter $J_{v}/J_{0} = -0.62$, somewhat larger than in $\rm CrI_{3}$~\cite{2021Magnetoelectric}. 
The $J_{0}$ value includes all distant neighbor exchange coupling. 
We have plotted in Fig.~\ref{fig:temperature-magntoelectroni}(c) the ratio of interlayer and intralayer exchange coupling from $-$0.2 to $-$1.0 to account for the out-of-plane strain effect in the magnetoelectric response.

\section{Summary and conclusions}\label{sec:conclusion}
We have investigated the magnetic and magnetoelectric properties of monolayer and bilayer 1T-CrTe$_2$, clarifying the microscopic mechanisms that give rise to ferromagnetic and antiferromagnetic exchange interactions and their tunability under external perturbations. Using ab initio DFT calculations together with model analyses and Monte Carlo simulations, we showed how orbital hybridization, exchange pathways, and spin–orbit coupling determine the ground states and critical temperatures of this layered dichalcogenide.

Monolayers were found to host intralayer ferromagnetism mediated by superexchange via Te ligands, while bilayers favor interlayer antiferromagnetism stabilized by super-super-exchange across the van der Waals gap. Analysis of the competing $t_{2g}$–$t_{2g}$ and $t_{2g}$–$e_{g}$ hopping channels provides a microscopic rationale for these phases and complements conventional total-energy approaches. We further demonstrated that the magnetocrystalline anisotropy energy (MAE) is highly tunable: changes in Hubbard U, strain, or interlayer spacing can drive a crossover between in-plane and out-of-plane easy axes.

Our Monte Carlo simulations predict Curie and Néel temperatures close to or above room temperature, and show that bilayers exhibit a sizable magnetoelectric response under perpendicular electric fields, despite the ground state being robust against large applied fields. These findings highlight 1T-CrTe$_2$ as a rare 2D magnet combining high transition temperatures with versatile tunability by strain, correlations, and electric fields.

While quantitative values remain sensitive to computational parameters and real samples may deviate from ideal AA stacking, the overall trends point to broad opportunities for 2D materials spintronics. The ability to reversibly modulate anisotropy and interlayer coupling suggests practical routes toward gate-controlled spin states, non-volatile memories, and magnetoelectric logic devices. In sum, 1T-CrTe$_2$ emerges as a promising platform for spintronics at or near room temperature and a model system for exploring tunable magnetism in two dimensions.

\begin{acknowledgments}
This work was supported by Basic Study and Interdisciplinary R\&D Foundation Fund of the University of Seoul (2025) for F.L. and J.J., and by SERB grant no. SRG/2022/001102 and ‘IISER Kolkata Start-up-Grant’ Ref.No.IISER-K/DoRD/SUG/BC/2021-22/376. for  B.L.C.
We acknowledge computer time allocations from the Texas Advanced Computing Center, from KISTI through grant KSC-2020-CRE-0072, the resources of Urban Big data and AI Institute (UBAI) at the University of Seoul and the network support from KREONET.
\end{acknowledgments}

\bibliography{aps}
\end{document}